\documentclass[aps,amsfonts,amsmath,prd,preprint,tightenlines,showpacs,nofootinbib]{revtex4}

\def\be{\begin{equation}}
\def\ee{\end{equation}}
\def\bea{\begin{eqnarray}}
\def\eea{\end{eqnarray}}
\def\bml{\begin{subequations}}
\def\blea{\bml\begin{eqnarray}}
\def\elea{\end{eqnarray}\end{subequations}}
\usepackage{epsfig}

\def\mpl{m_{\text{pl}}}
\def\Teq{T_{\text{eq}}}
\def\Tee{T_{\text{ee}}}
\def\teq{t_{\text{eq}}}
\def\tee{t_{\text{ee}}}
\def\tnecklace{t_{\text{necklace}}}
\def\GeV{\text{GeV}}

\begin{document}

\title{Monopole annihilation in cosmic necklaces}

\author{Jose J. Blanco-Pillado}
\email{jose@cosmos.phy.tufts.edu.edu}
\author{Ken D. Olum}
\email{kdo@cosmos.phy.tufts.edu}
\affiliation{Institute of
Cosmology, Department of Physics and Astronomy, Tufts University,
Medford, MA  02155}

\begin{abstract}

A sequence of two symmetry breaking transitions in the early universe
may produce monopoles whose flux is confined into two strings each,
which thus assemble into ``necklaces'' with monopoles as beads. Such
``cosmic necklaces'' have been proposed as a source of
ultra-high-energy cosmic rays. We analyze the evolution of these
systems and show that essentially all monopoles annihilate or leave
the string at early times, after which cosmic necklaces evolve in a
similar way to a network of ordinary cosmic strings. We investigate
several modifications to the basic picture, but in nearly all cases we
find that too few monopoles remain on the necklaces to produce any
observable cosmic rays.  There may be a small window for 
superconducting condensates to prevent annihilations, but only if both
the string and the condensate scale are very high.

\end{abstract}


\maketitle


\section{Introduction}

Conventional acceleration mechanisms within known astrophysical
objects face serious difficulties in accounting for the observation of
the ultra-high energy cosmic rays (UHECR's) with energies above
$10^{20}$ eV \cite{FlysEye}.  This has motivated the search for
alternative scenarios that could explain the origin of such stupendous
energies.

Topological defects (for a review see \cite{VS-book,HK-book})
are in this regard a very natural candidate to produce high energy 
radiation that could explain the UHECR observations. On the one hand, 
they are created during a phase transition in the early universe and 
are therefore associated with an energy scale typically much higher 
than $10^{20}$ eV. On the other hand, due to their topological stability
they can survive for long periods of time, allowing them to release
this energy today, when the universe is much colder and their
energies are very atypical. Several such scenarios have been proposed 
in the literature based on different kinds of defects such as
monopole-antimonopole pairs, cosmic strings, superconducting strings
or hybrid defects. (See \cite{BS,BBV} and references therein).

In this paper we will consider in detail the model proposed in
\cite{BV} based on the hybrid defects known as ``cosmic
necklaces''. This type of defects can be produced due to a sequence of
symmetry breaking phase transitions with either of the following structures,
\blea
G &\rightarrow& U(1) \rightarrow Z_2\label{eqn:g}~,\\
G &\rightarrow& H \times U(1) \rightarrow H \times Z_2\label{eqn:gh}~,
\elea where $G$ is a semi-simple group. In both of these scenarios, 
the first transition produces monopoles of mass $M \sim \eta_M/e$, 
where $\eta_M$ characterizes the energy scale of the universe at that
time. The second transition at a lower scale $\eta_s$ gives rise to 
cosmic strings of energy per unit length given 
by\footnote{We use units in which $c =\hbar= k_B = 1$ throughout.} 
$\mu \sim \eta_s^2$.

Each string confines half of the $U(1)$ magnetic flux coming from a
monopole.\footnote{Thus this $U(1)$ cannot be the usual
  electromagnetism, but some different field whose flux is confined.
  In the case of Eq.\ (\ref{eqn:gh}), the monopole may also be the
  source of other fluxes that are not confined.} This leads to a
situation in which each monopole is attached to two strings, each of
which has an antimonopole on the other end. This process creates a
network of infinite strings and loops with monopoles spaced along the
strings\footnote{We assume here that the monopoles and antimonopoles
  are randomly located.  If instead the particles are in nearby pairs,
  for example as in Ref.~\cite{Khlopov}, the two strings from a given
  monopole would often go to the same antimonopole, leading to
  two-particle loops similar to the strings discussed in
  Ref.~\cite{BPO}, with few long necklaces.}, similarly to beads on a
necklace, giving rise to the name ``cosmic necklaces''.  Cosmic
necklaces have been considered an attractive source for
ultra-high-energy cosmic rays \cite{BV}, because the monopoles could
annihilate at late times to produce high energy particles.\footnote{In
  the case of Eq.\ (\ref{eqn:g}), we imagine that there is some
  very-high-energy connection between the group G and the standard
  model, which permits monopoles to decay into ordinary particles.}

Particle physics models with this type of symmetry breaking pattern
have been previously discussed in the literature \cite{HK,AE}. Similar
objects, although with somewhat different properties, have also been
discussed more recently within the string theory context
\cite{Matsuda,DFG,LW,Dasgupta,Ben,Lake}.

The cosmological evolution of monopoles depends on the distribution of
energy between the monopoles and the strings.  The essential parameter
\cite{BV} is the ratio of the energy in monopoles to that in string
segments,
\be
r = \frac{M}{\mu d}~,
\ee where $d$ is the average distance between monopoles.  If $r\ll 1$,
the string evolves essentially without regard to the monopoles.  The
monopoles in such a necklace move along the string like beads on a
moving wire without significant back reaction on the motion of the
string itself.

On the other hand, if $r$ is significant the motion is different from
an ordinary cosmic string without monopoles.  The typical velocity is
 \cite{BV}
\be
v \sim 1/\sqrt{1 + r}\,.
\label{v}
\ee
If $r\gg1$, then $v \sim 1/\sqrt{r}$, which one can verify by treating the
necklace as a nonrelativistic string with tension $\mu$ but mass
per unit length $r\mu$.

Ref.~\cite{BV} studied the evolution of the parameter $r$ neglecting
the effect of monopole annihilation and concluded that the system
may evolve towards a state with $r \gg 1$.  On the other hand,
Ref.~\cite{SMO} studied the dynamical evolution of loops with $r \ll
1$ and found that in this case monopole annihilation is very
efficient. This is not very surprising, taking into account that in
this limit, the monopoles have no significant effect on the system and
they are therefore thrown around by the motion of the string. This
process gives a longitudinal velocity to the monopoles on the strings
close to the characteristic transverse string velocity, which in this
case is relativistic \cite{SMO}.

The parameter space of models described in \cite{BV} is very wide due
to the vast variation of possible scales allowed for the monopole mass
$M$ (whose only restriction is to be above the highest energy observed
in the UHECR spectrum, $M > 3\times 10^{11} \GeV$) as well as the string
tension $\mu$. On the other hand, it is clear that the necklace
network should be able to release enough energy to explain the UHECR
and therefore we should ensure that the fraction of the critical
density in the monopoles on the necklaces is larger than that in
UHECR,
\be
\Omega_M > \Omega_{\text{UHECR}}~.
\ee
For a scaling distribution of cosmic strings moving relativistically,
we expect $\Omega \sim G\mu = \mu/\mpl^2$ \cite{VS-book}.  For a network
moving with speed $v$, all scales on the string network will be larger
at a given time by a factor $1/v$ \cite{BV}, so the contribution from
strings is $\Omega\sim G\mu/v^2$.  The mass in monopoles is $r$ times
that in strings, so
\be\label{Omegadef}
\Omega_M \sim \frac{r(1+r)\mu}{\mpl^2}\,.
\ee
To match the observed flux of UHECR with the model of Ref.~\cite{BV}
implies $\Omega_{\text{UHECR}} \sim 10^{-10}$, so we require
\be\label{Omegabound}
\Omega_M \gtrsim 10^{-10}
\ee
for a viable model.

In this paper, we show that monopole annihilation in cosmic necklaces
is in fact a very efficient process to the point that we have not been
able to find any region of the parameter space in simple models that
satisfies the above requirement. Monopole-antimonopole interactions
rapidly annihilate monopoles or break off monopole-antimonopole
pairs. The remaining string behaves as a normal cosmic string network
while the pairs evolve as described in~\cite{BPO}.

The plan of this paper is as follows.  We first consider the simple
case of smooth, non-superconducting, Type II (or BPS) strings.  In
section \ref{sec:velocity} we show that monopoles acquire a typical
velocity along the strings similar to the transverse string velocity
$v$.  In section \ref{sec:interaction}, we investigate what happens
when a monopole and an antimonopole come together along the string.
We discuss the probabilities for annihilation, reflection, and expulsion
of the pair from the string.  In section \ref{sec:result} we calculate the
effects of these processes and show that the monopole density rapidly
decreases.

In section \ref{sec:fluc} we study the effect of correlations and
statistical fluctuations in monopole velocities on the conclusions of
this paper.  In section \ref{sec:wiggles}, we consider how wiggliness
affects the monopole motion.  In \ref{sec:unconfined}, we consider the
case of monopoles which have some fluxes that are not confined, in
addition to the flux that forms the strings.  In section
\ref{sec:typeI} we consider type I strings, which attract each other
and so make the detachment process less common. In all these cases, we find
our conclusions about rapid annihilation unchanged.  In section
\ref{sec:superconducting}, we consider the case of strings with
superconducting condensates, which can oppose the motion of the
monopoles along the string.  In this case we find that if the
condensate particles are trapped between monopoles, there is a small
window of parameter space in which annihilations are prevented and so
there may be the possibility of a viable model.  We conclude in section
\ref{sec:discussion}.


\section{Monopole motion}\label{sec:velocity}

If the two strings attached to a monopole point in opposite
directions, there is no net force on the monopole.  The monopole can
move freely along the string, reducing the length of the string on
one side and increasing the length on string on the other.  In the
limit $r\ll 1$, the back reaction of the monopole on the string is
small, and we can treat the monopoles as beads sliding freely along a
moving wire.

How rapidly should the typical monopole move along the string?  In
certain very simple cases, such as a string with standing waves and
monopoles at the nodes or antinodes, the monopoles will not move.  But
in general, the motion of the string can give longitudinal velocities
to the monopoles.

As a simple example, consider a $r\ll 1$ necklace where the string is
made up of a series of kinks and the motions of the different segments
are uncorrelated.  Suppose a monopole has longitudinal velocity
$v_\parallel$ along a string segment moving with velocity $v$.  Since the
monopole is constrained to move with the string, its transverse
velocity is also $v$, and its overall velocity is given by $v_M^2 =
v^2+v_\parallel^2$.

Now suppose the monopole suddenly finds itself on a different string
segment making angle $\theta$ with the overall direction of monopole
motion.  The monopole now has a new longitudinal velocity $v_\parallel' =
v_M \cos\theta$ so $v_\parallel'^2 = \cos^2\theta(v^2+v_\parallel^2)$.

Let us average over all realizations and look for a stationary mean
squared monopole velocity.  We find
$\langle v_\parallel'^2\rangle = \langle\cos^2\theta\rangle(\langle
v^2\rangle+\langle v_\parallel^2\rangle)$.  By assumption the old and
new string directions are uncorrelated, so
$\langle\cos^2\theta\rangle= 1/3$, and $\langle v_\parallel'^2\rangle = \langle
v^2\rangle/2$.  We will not be concerned with the factor of 2, only with
the fact that the longitudinal monopole velocity is comparable to the
string velocity.

A realistic string would be much more complicated than the simple
model above.  There would be not only sharp kinks but also smooth
variations of direction and velocity.  It would also be possible for
energy to flow back and forth between the string and the monopole.  In
this case, we can make an argument of equipartition.  The average
energy in the string between monopoles is $\mu d$.  The energy of a
monopole with velocity $v_M$ is $M v_M^2/2$ in the nonrelativistic
case.  For these to be roughly equal, we should have $v_M$ of order
$\sqrt{\mu d/M} = 1/\sqrt{r}$.  The monopole kinetic energy should be
divided between longitudinal and transverse motion, both of which
should then have velocities of order $1/\sqrt{r}$.  Thus once again we
conclude that the longitudinal monopole velocity is of order
$1/\sqrt{r}$ if $r\gg 1$.

The motions of adjacent monopoles are not independent.  The string
will be smooth on some correlation length $\xi$.  Monopoles within
distance $\xi$ will be subjected to similar motions and will thus
tend to acquire similar longitudinal velocities.  As a simple model,
we can take monopoles to consist of trains of order $\xi/d$ monopoles
all moving similarly with random longitudinal velocities of order $v$.


\section{Monopole interaction}\label{sec:interaction}

As discussed in the previous section, the monopoles will acquire
velocities along the string.   Suppose a monopole and an
antimonopole are moving toward each other.  Unless some process causes
them to turn around again (which will be the subject of later
sections), the monopole and antimonopole will come together.  What
will happen then?

First suppose that the string is straight and the monopole velocity is
entirely along the string.  If the monopole and antimonopole collide
head-on, they will annihilate.  The monopole mass-energy will then be
radiated in high-energy particles and lost to the necklace.

\begin{figure}
\centering\leavevmode
\epsfysize=8cm \epsfbox{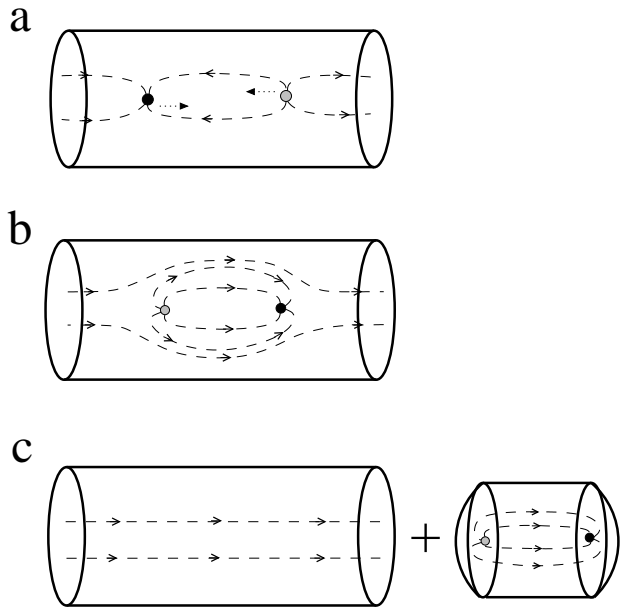}
\caption{Monopole (gray) and antimonopole (black) passing by inside a
  string.  Dashed lines with arrows show the magnetic flux lines.  First
  (a) the particles are moving towards each other inside the string.
  After they pass there are 3 strings worth of flux in the
  region between the monopoles.  By rearranging flux lines (b), all
  flux from the monopole goes to the antimonopole, so the
  monopole-antimonopole pair can detach itself from the rest of the
  string (c).}
\label{FIG1}
\end{figure}

However, if the string scale and the monopole scale are well
separated, the monopoles might easily pass by each other inside the
string, as shown in Fig~\ref{FIG1}(a)--(b).
There is now an string with winding number $n=3$ between the monopoles
and $n=1$ strings outside them.  They thus feel a net force pulling
them back toward each other.  In the critically coupled (BPS) case,
the force is $2\mu$.  For Type II strings it is even larger.  The Type
I case will be discussed in Sec.~\ref{sec:typeI}.

In an ideal geometry, this force would eventually reverse the
monopole velocities so that they pass back past each other.  However,
during the time that the monopoles have passed, there is no longer any
topological constraint holding them to the string, and it is possible
for them to detach as shown in Fig~\ref{FIG1}(c).

If the string is Type II, a string with winding number $n=2$ and
another with $n=1$ have less energy than a single $n = 3$ string.  It
is therefore favored for the monopole pair to detach itself from the
rest of the string, forming an $n=2$ string (or two $n=1$ strings)
containing all the flux between monopole and antimonopole.  If instead
the string has critical coupling, there is no energetic
advantage or disadvantage to detachment.  If the string is not exactly
straight, however, the monopoles have some transverse velocity which
will encourage detachment.  If the strings are Type I, detachment is
energetically disfavored.  Such strings will be the subject
of section \ref{sec:typeI}.

If the monopoles do not detach themselves from the string, the extra
tension of the $n=3$ string will eventually pull them back past each
other.  This gives another chance at annihilation, but if that fails
the net effect will be that the monopole and antimonopole bounce and
recede from each other.

\begin{figure}
\centering\leavevmode
\epsfysize=4cm \epsfbox{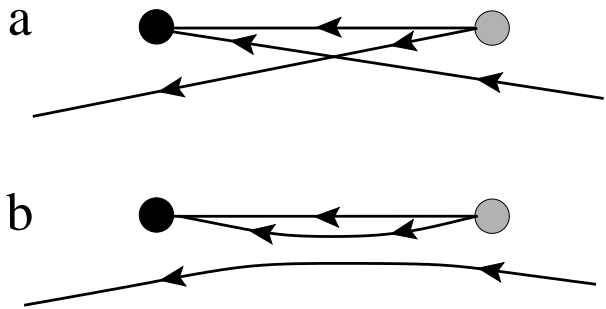}
\caption{A monopole and an antimonopole can pass each other to form a
  zigzag string (a). Some of the time there will then be a
  reconnection (b) which allows the monopole-antimonopole pair to detach
  itself from the rest of the string.}
\label{FIG2}
\end{figure}

If the monopole velocities are significantly misaligned, the minimum
distance of approach will be larger than the string radius.  In that
case the monopoles will pass outside the string and the result will be
a ``zigzag'' string, as shown in Fig.~\ref{FIG2}.
There is now a nontrivial chance that the strings will cross and
reconnect.  Then, as above, the monopoles are detached into a
2-monopole loop, leaving the rest of the string behind. This is
somewhat similar to the situation described in \cite{SMO}, where
non-periodic oscillations of a loop led to frequent
intercommutations.  If this does not happen, each monopole feels a
force $2\mu$ pulling it back in the direction from which it came, so
eventually monopole and antimonopole velocities will be reversed and
again monopoles will have reflected.

We will describe the result of a close approach between monopole and
antimonopole with a probability $P_a$ for annihilation, $P_d$ for
detachment, and $P_r = 1-P_a-P_d$ for reflection.  For most purposes,
we do not need to distinguish between annihilation and detachment, so
we write $P_x = P_a+P_d$ for the probability of any process that
removes the monopoles from the string.


\section{Resulting monopole density}\label{sec:result}

In the previous section we argued that (at least in the simple case of
smooth, non-superconducting, non-type-I strings), the probability of
annihilation or detachment, $P_x$, is significant. Let us now
estimate how this affects the value of $r$.

At very early times, the network of necklaces is dominated by
friction.  Necklaces therefore do not move even with velocity 
$v$ so the monopoles are not significantly annihilated
(although, even in this case, $r$ is decreased by the expansion of the
universe).  As we discuss in the Appendix, this era ends at a
time $\tnecklace$ at which it becomes possible for the necklaces to
move at the characteristic velocity $v\sim1/\sqrt{1+r}$.  If $r$ is
significant, $\tnecklace$ is earlier than the usual frictional
timescale $t_*$ where strings without monopoles can move
relativistically, because of the slower characteristic velocity of
necklaces.

After $\tnecklace$, monopoles will acquire velocities along the string
of order $v$, and so annihilate or detach if $P_x$ is large. In the
absence of friction this process increases the monopole distance to $d \sim v t$.  Particles
further away than this have had no opportunity to encounter each
other.  As $r$ declines, $v$ increases, and thus $r$ decreases at 
least as $t^{-1}$.  In the radiation era, Eq.~(\ref{tneckr}), decreasing $r$ increases $\tnecklace$
proportionately to $r^{-2}$.  Since this grows at least as fast than
$t$, friction limits the growth of $v$ and the necklaces stay at the
edge of the frictional regime.\footnote{In fact annihilations take
  place even when the necklace speed is restricted by friction, but we
  have ignored those.  Taking them into account would make the effect
  of monopoles even less important.}  In the matter era,
Eq.~(\ref{tneckm}), decreasing $r$ increases $\tnecklace$
proportionately to $r^{-1}$, so $\tnecklace$ keeps pace with $t$.
Thus in either case we stay at the edge of the frictional regime,
until we arrive at $t_*$ with $r\sim 1$.  Except for very light
strings, $t_*$ is well earlier than the age of the universe today,
$t_0$.  Since $r\lesssim 1$ at $t\gtrsim t_*$, $v$ is of order 1 from
$t_*$ until $t_0$.  (See Eq. \ref{v}.) Then by $t_0$ the distance
between monopoles increases to
\be
d \sim t_0 \sim 10^{41}\GeV^{-1}
\ee
and so
\be
r=\frac{M}{\mu d} \sim\frac{M}{\mu t_0}\ll 1\,.
\ee
This gives
\be
\Omega_M \sim \frac{M}{\mpl^2t_0}< 10^{-62}
\ee
so the effects of monopoles are completely negligible.

If the string scale is less than about
\be
\mu_* = \frac{\mpl^2}{t_0\Teq}\sim (10^3\GeV)^2\,,
\ee
$t_*$ is still in the future.  In that case, we can have $r
> 1$ today.  But we are still limited by the constraint that the
necklaces should be at the edge of the frictional regime.  Then from
Eq.\ (\ref{tneckm}),
\be
r\sim\frac{\mpl^2}{\mu t_0\Teq}
\ee
Since $r > 1$, Eq.\ (\ref{Omegadef}) gives
\be\label{eqn:lightOmega}
\Omega_M\sim\frac{\mpl^2}{\mu t_0^2 \Teq^2}<10^{-30}
\ee
if the lightest possible strings have energy scale 100 GeV.  This is
still 20 orders of magnitude too low to satisfy  Eq.\ (\ref{Omegabound})

In fact, this conclusion holds even if the probability of annihilation
or detachment is quite small.  If a monopole-antimonopole interaction
does not lead to annihilation or detachment from the string, the
particles effectively bounce.  After that they head away from each
other and have an opportunity to annihilate with their partners on
their other sides.  Thus in the case where $P_x$ is significantly less
than 1, monopoles and antimonopoles move back and forth with velocity
$v$ on segments of string of length $d$, and so have opportunities to
interact separated by a time $d/v$.  The timescale of annihilation or
detachment is then $t_x = d/(vP_x)$.  A value of $r$ that gives $t_x$
earlier than the present time would have decreased before now, so the
largest $r$ that can currently exist is found by setting $t_x = t_0$.
Using $d=M/(\mu r)$, $v\sim 1/\sqrt{1+r}$, we find
\be\label{rfromPx}
\frac{r}{\sqrt{1+r}} \sim \frac{M}{\mu t_0 P_x}
\ee
Taking extreme values of the parameters to make $r$ as large
as possible, namely  $M = 10^{17}\GeV$ and $\mu = (100\GeV)^2$,
and $t_0 \sim 10^{41}\GeV^{-1}$, we
find that the right hand side is at most $10^{-28} P_x^{-1}$.  Thus
providing $P_x> 10^{-28}$, $r<1$ and
\be
\Omega_M \sim \frac{M}{\mpl^2 t_0 P_x}<10^{-34}
\ee
Even smaller values of $P_x$ will be discussed in Sec.~\ref{sec:typeI}.

Thus we find that in the simple cases considered so far, monopole
annihilation or detachment reduces the number of monopoles on
necklaces to negligible values, so that they cannot act as a source of
ultra-high-energy cosmic rays.  In the sections below, we will discuss
several different modifications of this picture.


\section{Correlations and fluctuations of monopole motion}\label{sec:fluc}

As described in section \ref{sec:velocity}, the velocities of adjacent
monopoles will be correlated, so that there are trains with
alternating monopoles and antimonopoles all moving with similar
velocities.  Particles within a train will not approach at velocity
$v$, as we assumed before, so the above analysis requires
modification.  On the other hand, groups separated by larger than the
correlation length $\xi$ will be moving independently.  Suppose two
groups of length $\xi$ and having $\xi/d$ monopoles each are moving
together, each at speed $v$.  If $P_x\ll 1$, when the leading member
of one group reaches the other, it will reflect.  It will then move
back toward its own group, reflect off the second member, and very
soon the coherent motion will be destroyed.  The monopoles will now be
bouncing back and forth as above, and the analysis of
Sec.~\ref{sec:result} will apply.

On the other hand, if $P_x\approx 1$, the leading members of the
groups will annihilate with each other, then the subsequent members in
turn will do likewise.  In time $\xi/v$ the trailing members of the
two groups will reach each other and annihilate, and both groups will
be gone from the string.  Let us see if the existence of correlations
changes our previous conclusions in this case.

We will consider the following simple model.  Suppose that the string
is split up into uncorrelated sections of length $\xi$ that are moving
either to the right or to the left.  This understates the annihilation
rate, because in the real situation there are no precise trains moving
in lockstep.  Any given section will have the section ahead coming
toward it on average half the time, in which case it will annihilate
with that section in time $\tau = \xi/v$.  Thus in time $\tau$, the
number of monopoles will decrease by a factor of 2, and so the average
monopole distance $d$ will increase by that factor.  We will
approximate that in the same time, the motion of the string
reorganizes the monopoles into sections with a new, larger value of
$\xi$ and redistributes the monopoles and empty regions, so that we
can once again treat the monopoles as uniformly distributed.  The
result is that $d$ doubles in time $\tau$.

Now, how large can $\xi$ be?  From causality, the correlation length
on the string cannot exceed $vt$.  But it takes time for monopoles
to forget their original velocities and acquire new ones from the
moving string.  So $\xi$ must be less than $vt$, and thus $\tau<t$.
Consequently, when $t$ doubles, $d$ increases by more than a factor of
2.  Thus $r$ falls more quickly than $1/t$.

This process begins when friction no longer prevents necklaces
from achieving velocity $v$, at time $\tnecklace$.  According to the
analysis above, $r$ would start to drop faster than as $1/t$, which
means that monopole density tracks the edge of the frictional regime,
as described in Sec.~\ref{sec:result}, so $r\sim 1$ at $t=t_*$.

For light strings, this is sufficient to show that necklaces cannot be
an observable source of cosmic rays, as follows.  If $r\sim 1$ at
$t=t_*$, then by Eq.~(\ref{Omegadef}), $\Omega_M \sim \mu/\mpl^2$ then.
After the initial symmetry breaking transition, there is no process
that creates monopoles, so their number in a comoving region cannot
change, and so $\Omega_M$ cannot increase from the time of
matter-radiation equality, $\teq$, until today.  In fact, further
annihilations will decrease $\Omega_M$, but we don't even need to
consider that.  In order to have $t_*\agt\teq\sim \mpl/\Teq^2$ we
require $\mu\alt\mpl\Teq\sim10^{10}\GeV^2$, so $\Omega_M \alt
\Teq/\mpl \sim 10^{-28}$, which is negligible.  If string are lighter
still, so that $t_*$ is in the future, then as long as we stay at
the edge of the frictional regime, we have the unobservable case of
Eq.\ (\ref{eqn:lightOmega}).

If the string scale is larger, then $t_*$ will be in the radiation
era.  In that case, monopoles redshift like matter in the radiation
era, so $\Omega_M$ grows proportionally to the scale factor, which in
turn is proportional to $t^{1/2}$.  Thus, again ignoring annihilations, we
find that today
\be
\Omega_M \alt \frac{\mu}{\mpl^2}\left(\frac{\teq}{t_*}\right)^{1/2}
\sim \frac{\mu^2\teq^{1/2}}{\mpl^{7/2}}
\ee
To agree with Eq.~(\ref{Omegabound}) requires
\be
\mu^2 \agt 10^{-10} \mpl^{7/2} \teq^{-1/2}
\ee
or
\be\label{mubound}
\mu \agt 10^{19}\GeV^2
\ee
If $\mu$ is too small to obey Eq.~(\ref{mubound}), then as long as the
monopole density tracks the frictional regime until $t = t_*$, there
will be no observable cosmic rays.

For strings obeying Eq.~(\ref{mubound}), we note that after $t_*$, $r$
continues to drop as $1/t$, so today we have
\be
r\sim \frac{t_*}{t_0}= \frac{\mpl^3}{t_0\mu^2}\,,
\ee
and 
\be
\Omega_M \sim \frac{\mpl}{t_0\mu} < \sim  10^{-41}\,,
\ee
once again unobservably small.

One might also be concerned about the effect of statistical
fluctuations on the number of monopoles moving in the two directions.
Consider a long segment of string containing $N$ trains of monopoles.
By chance we expect there to be of order $\sqrt{N}$ more trains moving
in one direction than another.  If the particle velocities were fixed
over time, the left-movers and right-movers would annihilate with each
other, leaving about $\sqrt{N}$ trains moving in the majority
direction.  However, in our case, the monopole motion is not fixed.
Instead, as time passes, the motion of the string will affect the
monopole motions in such a way that any time there will be roughly the
same number going to two directions, rather than preserving any
preexisting asymmetry.  This prevents asymmetries from growing as a
fraction of the total monopole density, and thus the effect described
here is minimal.


\section{Wiggly strings}\label{sec:wiggles}

A realistic string is not exactly straight.  Instead, it has some
spectrum of ``wiggles''.  As monopole and antimonopole approach each
other, the wiggles can be compressed between the two monopoles,
similarly to photons between two mirrors, and contribute an effective
repulsion.  Could such a force prevent the monopoles from colliding?
We will show that very large wiggles might be barely
enough to prevent collisions, but that in fact one process or
another will always damp out the wiggles, so that a cosmologically
significant number of monopoles will never remain.

\subsection{Force due to wiggles}\label{sec:wiggleforce}

\begin{figure}
\centering\leavevmode
\epsfysize=4cm \epsfbox{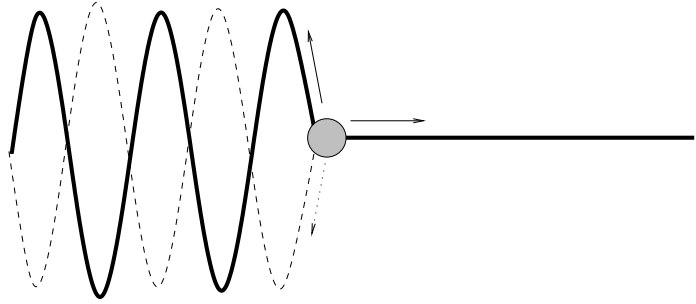}
\caption{Force on a monopole due to wiggles.  The monopole is moving
  to the left, leaving a straight string to its right and a very
  wiggly string to its left.  The straight string always exerts a
  force $\mu$ to the right.  The wiggly string exerts an instantaneous
  force $\mu$ in a varying direction.  For a sufficiently wiggly string
  the average of the varying force will be small, leaving a net rightward
  force $\mu$.}
\label{FIG3}
\end{figure}

First let us analyze the repulsive effect of wiggles.  We expect that
after the monopole has moved in a given direction, it has a very
wiggly string on one side and a much straighter string on the other
side.  The effective tension of the wiggly string could be as low as
zero \cite{VS-book}, while the effective tension on the straight side
could be as high as $\mu$.  One might think that the wiggles could
exert a true repulsion that would add to the force on the straight
string side of the monopole, but this is not correct.  In a close-up
picture of the monopole (e.g., Fig.~\ref{FIG3}),
we see that the string moves around but always exerts an instantaneous
force of $\mu$ in the direction in which it is pointing.  If the set
of directions of the string to the left of the monopole are uniformly
distributed, this force could average to zero but even if this is not
the case we would certainly not expect it to point in the direction
opposite to the direction from one monopole to the other.

Now consider two monopoles moving toward one another.  They will
typically be separated by distance $d$.  As we explained above the
maximum value of the repulsive force is $\mu$ and can thus remove at
most an amount of kinetic energy of the order of $\mu d = M/r$ as the
monopoles come together.  The typical velocity is $v$, so the total
kinetic energy is $M v^2 \sim M/(1+r)$.  If $r\agt 1$, these are about
equal, so the only most optimistic effective repulsion has the right
size to bring the monopoles to a halt.  On the other hand, if $r \ll
1$, then the repulsive force of wiggles can easily be sufficient.  We
will see in the following, however, that there are always processes
that damp the wiggles, and so significant repulsive force is not
available in situations of relevance to cosmic rays.

The same conclusions apply if the monopoles move in groups.  Suppose
there are two groups of length $\xi$, so each group contains $N =
\xi/d$ monopoles and thus $N$ times more kinetic energy than in the
single particle case.  If the distance between group centers is also
$\xi$, then the monopole kinetic energy that can be removed by the
repulsive force of wiggles is also larger than the single particle
case by factor $N$, so the comparison is just as before.

\subsection{Damping of wiggles}

Even if wiggles could in principle keep the monopoles from colliding,
they would have to survive long enough to do so.  Wiggles can be
damped by friction or by gravitational back reaction, and we will see
that one process or the other prevents there from being sufficient
wiggliness to keep monopoles from colliding over a long period of the
history of the universe.

First consider the frictional damping of wiggles on the string between
a pair of monopoles.  In the Appendix we see that wiggles on the
string will be exponentially damped until the universe reaches the
time $t_*$ when a string with a given $\mu$ would no longer be
friction dominated.  Thus wiggles do not affect string motion until
$t_*$, and so, by the argument of Sec.~\ref{sec:fluc}, only strings
obeying Eq.~(\ref{mubound}) are of interest.

After $t_*$, frictional damping is ineffective, but now the strings
are moving relativistically, and so they can be damped by the emission
of gravity waves.  Consider wiggles of wavelength $\lambda$ and
amplitude $\epsilon\lambda$ on the string segment between two
monopoles.  The parameter $\epsilon$ is a dimensionless measure of the
strength of the wiggle, and if $\epsilon\ll0$, the wiggles increase
the mass density and decrease the tension of the string by factor
$1+\epsilon^2/4$.  Such wiggles emit gravity waves with power per unit
length\footnote{We do not need to consider the effect discussed in
  Ref.~\cite{Siemens:2001dx} in which radiation is suppressed if
  wiggles of different sizes are moving in the two directions.  In
  order to exert a repulsion the wiggles must reflect off the
  monopoles and so move evenly in both
  directions.}\cite{Hindmarsh:1990xi,Siemens:2001dx}
$(\pi/4) G \mu^2 \epsilon^4/\lambda$.
The energy emitted comes from the wiggles, so
\be
\frac{d}{dt}\epsilon^2 = -\pi G \mu^2 \epsilon^4/\lambda\,,
\ee
which has the solution
\be
\epsilon^2 = \frac{\lambda}{\pi G \mu (t + \text{const})}
\ee
If the wiggles had amplitude $\epsilon_0$ at some initial time, then
an interval $\Delta t$ later the amplitude is
\be
\epsilon = \left(\frac{\lambda}{\pi G \mu \Delta t + \lambda/\epsilon_0^2}\right)^{1/2}
\ee

Now repulsion can only result from wiggles whose wavelength obeys
\be\label{eqn:lambda}
\lambda\mu \lesssim M\,.
\ee
Wiggles longer than this will travel past a monopole rather than
being reflected, and so will not lead to a force on the monopole.
Thus regardless of $\epsilon_0$, we find
\be
\epsilon\alt\left(\frac{M}{\pi G \mu^2 \Delta t}\right)^{1/2}
\ee
for relevant wiggles.

As argued above, we only need to consider $\mu\agt 10^{19}\GeV^2$.
With $\Delta t$ the present age of the universe, we find $\epsilon\alt
10^{-11}$, so primordial wiggles have been damped into insignificance
by gravitational wave emission.  In fact, after even the much shorter
time $\Delta t = t_*$, we already have $\epsilon \alt \sqrt{M/\mpl} \alt
10^{-1}$.  By the argument of Sec.~\ref{sec:wiggleforce} such wiggles
are too small to prevent monopoles from colliding if $r > 1$.

Of course the wiggles might not be primordial.  If wiggles are
produced by intercommutations, then the timescale for production will
be of order the Hubble time, since the distance between strings is of
that order.  Then the previous calculation applies and $\epsilon$ is
trivial.  Suppose instead that wiggles are produced by the
oscillations of loops \cite{SMO}.  The smallest loop that can persist
until the present day has size $l \sim G\mu t_0$.  While such a loop
might be producing wiggles as it oscillates, in one oscillation the
wiggles will be damped to level $\epsilon\sim \sqrt{M/(G^2\mu^3 t_0)}
\sim 10^{-1}$.  These wiggles will have no effect unless $r<1$.  But
in that case, $\Omega_M$ is too small to give observable effects unless
$G\mu\agt 10^{-10}$, in which case $\epsilon \alt 10^{-16}$.

Thus, repulsion due to wiggles can never revive necklaces as a source
of observable cosmic rays, regardless of the string tension.

\section{Monopoles with unconfined fluxes}\label{sec:unconfined}

If monopoles have an unconfined flux in addition to the flux carried
by the strings, and there is a plasma of particles charged with
respect to that flux, then a monopole will experience a frictional
force \cite{VS-book} of order $T^2 v$, which results in power
dissipation of order $T^2 v^2$.  Comparing this with the monopole
kinetic energy of order $M v^2$ yields a damping time $M/T^2$, which
is always much less than the Hubble time, $\mpl/T^2$.  Thus as long as
there is a background of charged particles, the monopole motion will
be friction-dominated.  In this case, the monopoles start to acquire
significant velocities only when the charged particles annihilate, at
the latest at the time of electron-positron annihilation, $\tee$.

In the standard scenario of Sec.~\ref{sec:result}, the monopoles begin
annihilating at $\tee$, and the results of that section still apply.
If instead the annihilation rate is reduced by the considerations of
Sec.~\ref{sec:fluc}, the monopole density falls as $1/t$.  But even
this is enough to prevent an observable signal.  At the time of
nucleosynthesis, which is essentially the same as $\tee$, we require
$\Omega_M \alt 1$ to preserve the correspondence with observation of
light element abundances.\footnote{This is simply the constraint that
  the universe should be radiation dominated during nucleosynthesis.  A
  much stronger bound can be derived by considering effects of energetic
  particles released by monopole annihilation during nucleosynthesis.}
Between $\tee$ and $\teq$, annihilations decrease the number of
monopoles in a comoving volume by factor $\teq/\tee \sim
(\Tee/\Teq)^2$.  Meanwhile, redshifting of relativistic particles
decreases the energy in a comoving volume by factor $\Tee/\Teq$,
leading to a net decrease in $\Omega_M$ by factor $\Tee/\Teq \sim
10^6$.  After equality, $\Omega_M$ simply declines as $1/t$, producing
an additional decline of $t_0/\teq \sim 10^6$.  Thus today we have
$\Omega_M \alt 10^{-12}$ in disagreement with Eq.~(\ref{Omegabound}).


\section{Type I strings}\label{sec:typeI}

\begin{figure}
\centering\leavevmode
\epsfysize=5cm \epsfbox{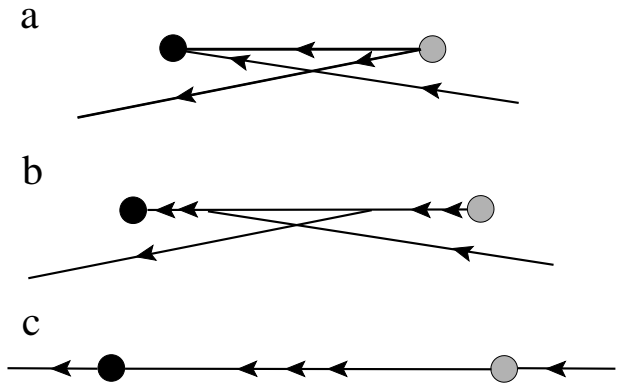}
\caption{If the monopoles are connected by Type I strings, regions
of higher winding will easily be formed in the collisions between monopole
and antimonopoles.}
\label{FIG4}
\end{figure}

One can also imagine that the strings that connect the
monopoles are Type I strings. These are characterized by an
attractive force between parallel static strings \cite{Rebbi}.  In
this case the energy density of an $n=3$ string is less than the sum
of an $n = 2$ string and $n = 1$ string.  This makes the detachment process
shown in Fig.~\ref{FIG1} energetically disfavored.  Furthermore, if
the strings form a ``zigzag'' string as in Fig.~\ref{FIG2}, it is
possible for them to form a ``zipper'' region where two
unit-winding strings become  attached to one another \cite{Laguna}.
Thus instead of the result shown in Fig.~\ref{FIG2}, we have that
shown in Fig.~\ref{FIG4}, in which the particles, after initially
passing outside the string nevertheless end up in a single string.

If the string is sufficiently Type I, perhaps all detachment processes
are suppressed, so that $P_d = 0$.  Instead of detachment, we end up
with the monopole and antimonopole moving away from each other inside
a single string.  The tension of the $n = 3$ string in between the
particles will eventually reverse their motion and bring them back
past each other.  In this case, there will be a probability for
annihilation on the return trip.  If this probability is large enough,
the monopole density will still be greatly decreased.

We can estimate the probability of annihilation as the geometrical
ratio between the size of the monopole to the string thickness,
\be\label{typeIPx}
P_x = P_a = \left(\frac{\eta_s}{\eta_M}\right)^2 = \frac{\mu}{M^2}
\ee
In anything but the most extreme cases, this gives $P_x > 10^{-28}$,
and thus by the argument of Sec.~\ref{sec:result}, $\Omega_M$ is negligible.
If we go to $M = 10^{17}$ GeV, $\mu = (100 \GeV)^2$,
with no probability for detachment,  Eq.~(\ref{typeIPx}) gives
$P_x\sim  10^{-30}$.  In such a case, Eq.~(\ref{rfromPx}) and
Eq.~(\ref{Omegadef}) give
\bea
r &\sim& \left(\frac{M}{\mu t_0 P_x}\right)^2\\
\Omega_M & < & \frac{M^4}{\mu^3 t_0^4 P_x^4 \mpl^2} = 
\frac{M^{12}}{\mu^7 t_0^4 \mpl^2}< 10^{-26}~.
\eea


\section{Superconducting strings}\label{sec:superconducting}

The strings formed in the second phase transition may have some
massless zero modes living on them, in other words they may be 
superconducting\footnote{Note that we use the
  terminology of superconducting to refer to any kind of persistent
  currents created by the zero modes localized on the string
  core and not only to the electromagnetically charged
  modes.} \cite{Witten}. These new degrees
of freedom change the effective equation of state of the string
and could have potentially important consequences for the motion
of the string itself \cite{VS-book}. Here we will be mostly interested
on the effect that these modes may have on the longitudinal velocity
of the monopoles. 

We can think of a superconducting string as a wire along which
massless charge carriers can flow.\footnote{Here we are thinking of
  fermionic superconducting zero modes, but bosonic superconductivity
  leads to similar limits.} These charge carriers act as a
one-dimensional gas and exert pressure that decreases the tension of
the string.  If the charge carriers can pass freely across a monopole,
then they have no effect on monopole motion along the
string.\footnote{Charge carriers are held to the string because they
  are massless inside but massive outside the string.  The existence
  of monopoles puts only a small gap in the string core, and it seems
  that charge carriers could easily tunnel across this gap.}  But if
the charge carriers are reflected by the monopole, they can be
concentrated between the monopole and antimonopole as they approach.
Reflection reverses the current but retains the charge of the carrier,
so the string between monopoles will have charge but no current.  The
charge density cannot grow without bound: eventually the charge
carriers can scatter and leave the string.  The maximum charge density
is dependent on the specific model.  If charge carriers can scatter to
produce light charged particles not bound to the string, the maximum
charge density will be low.  If not, it can be high enough to
substantially reduce the string tension, but it can never reduce it to
zero \cite{Peter1}.  As discussed in Sec.~\ref{sec:wiggleforce}, this
means that the effective repulsion is too small to prevent collisions
if $r>1$, but might do so if $r<1$.  Thus there may be a small window
for annihilations to stop if the string is superconducting, the scale
of superconductivity is nearly the same as the string scale, and the
string acquires a significant charge density.  This process works only
if $r < 1$, so $\Omega_M \alt G\mu$, and thus it is only of interest
for heavy strings with $G\mu\agt 10^{-10}$.


\section{Discussion}\label{sec:discussion}

We have shown that the monopoles on a necklace acquire typical
velocities along the string large enough to lead to very frequent
interactions between monopoles and antimonopoles.  We have studied a
wide variety of cases (strings of Type I or II, superconducting
strings, strings with wiggles, monopoles with additional unconfined
fluxes), but found in almost every case that interactions lead, at
early times, to monopole-antimonopole annihilation or to the monopoles
leaving the string in monopole-antimonopole pairs.  In the former case
the necklaces just become ordinary cosmic strings.  In the latter
case, instead of necklaces we have a population of
monopole-antimonopole pairs, and the subsequent evolution is just as
we discussed in Ref.~\cite{BPO}.  In neither case does the possibility
of necklaces add anything to the set of possible ultra-high-energy
cosmic ray sources.

The exception to the above is heavy superconducting strings with heavy
condensate scale.  In this case it seems barely possible that the
superconducting charge carriers could prevent the monopoles from
reaching each other.  In order to make a viable model, the string
scale must be quite high, $G\mu\agt 10^{-10}$, the scale of the
condensate must also be quite high, and there must not be any process
which scatters condensate particles into lighter particles that are not
bound to the string.  Furthermore, the charge carriers must not be
able to pass or tunnel between the segments of string on the two sides
of a monopole, and there must be some mechanism by which the string
acquires a significant fraction of the maximum charge density.

\section*{Acknowledgments}

We thank Marco Kneipp, Xavier Siemens, Alex Vilenkin and Tony Weinbeck
for helpful conversations. K.D.O. was supported in part by the
National Science Foundation under grants 0457456 and 0855447.

\appendix

\section{Friction}
 
In this appendix, we compute the time at which friction is important
for segments of string between monopoles and the time at which
friction is important for the motion of the necklace as a whole.  If
$r$ is significant, friction loses its effect sooner for the necklace
as a whole, because its velocity in the absence of friction is lower.

The frictional force per unit length on a string with velocity $v_s$
interacting with a thermal background of temperature $T$ goes as
\cite{Everett}
\be
F \sim T^3 v_s~,
\ee
so the damping power per unit length is
\be
Fv_s \sim T^3 v_s^2~.
\ee
We neglect relativistic factors here.  String segments between
monopoles are never ultra-relativistic, so the contribution of
relativistic effects is at most of order 1.

Excitations of wavelength $\lambda$ and amplitude $A$ have energy per
unit length of order $\mu A^2/\lambda^2$.  They give rise to a string
velocity of order $A/\lambda$, so the frictional power per unit length
is $T^3A^2/\lambda^2$ and the damping timescale is \cite{VS-book}.
\be
\tau_d \sim \frac{\mu}{T^3}~.
\ee

Features on the string will be exponentially damped, as long as
$\tau_d$ is less than the age of the universe.  In the radiation era,
that is $t\sim \mpl/T^2$, so we find that friction domination for
string segments ends at
\be\label{eqn:t*}
t_* \sim \frac{\mpl^3}{\mu^2}\quad\text{(Radiation)}~.
\ee
Although we did the calculation in a slightly different way, this is
just time when friction stops being important for a string network
without monopoles \cite{VS-book}.

If friction lasts until the matter era, we have instead
$t\sim\mpl/\sqrt{T^3\Teq}$, so
\be\label{eqn:t*matter}
t_* \sim \frac{\mpl^2}{\mu\Teq}\quad\text{(Matter)}~.
\ee

Now let us consider the damping of the necklace as a whole.  The
overall motion has $v \sim 1/\sqrt{1+r}$ \cite{BV}, kinetic energy per
unit length of order $(1+r)\mu v^2\sim \mu$, and frictional power per unit
length of order $T^3/(1+r)$.  The damping time for this case is thus
\be
\tau_d \sim \frac{(1+r) \mu}{T^3}~,
\ee
and friction for the motion of the necklace as a whole ends at time
$\tnecklace$, where
\bea\label{tneckr}
\tnecklace &\sim& \frac{\mpl^3}{(1+r)^2 \mu^2}\quad\text{(Radiation)}\\
\label{tneckm}
\tnecklace &\sim& \frac{\mpl^2}{(1+r) \mu\Teq}\quad\text{(Matter)} ~.
\eea

\end{document}